\documentclass[a4paper]{article}

\usepackage{icrc2013}
\usepackage[english]{babel}

\usepackage{amssymb}

\title{Constraints on Lorentz Invariance Violation with \textit{Fermi}-LAT Observations of Gamma-Ray Bursts}

\shorttitle{Constraints on Lorentz Invariance Violation with \textit{Fermi}-LAT Observations of Gamma-Ray Bursts}

\authors{
C. Couturier$^{1}$,
V. Vasileiou$^{2}$,
A. Jacholkowska$^{1}$,
F. Piron$^{2}$,
J. Bolmont$^{1}$,
J. Granot$^{3}$,
F. W. Stecker$^{4}$,
J. Cohen-Tanugi$^{2}$,
F. Longo$^{5}$
}

\afiliations{
$^{1}$ LPNHE, Universit\'e Pierre et Marie Curie Paris 6, Universit\'e Denis Diderot Paris 7, CNRS/IN2P3 \\
$^{2}$ LUPM, Universit\'e Montpellier 2, CNRS/IN2P3 \\
$^{3}$ Department of Natural Sciences, The Open University of Israel \\
$^{4}$ Astrophysics Science Division, NASA/Goddard Space Flight Center and Department of Physics and Astronomy, University of California \\
$^{5}$ Istituto Nazionale di Fisica Nucleare, Sezione di Trieste and Dipartimento di Fisica, Universit\`a di Trieste

}

\email{camille.couturier@lpnhe.in2p3.fr}

\abstract{
Arrival times of photons from four bright GRBs observed by \textit{Fermi}-LAT have been analyzed in order to detect a possible Lorentz Invariance Violation (LIV), related to a non trivial dispersion of light in vacuum. No delays have been detected and strong limits on the Quantum Gravity (QG) energy scale are derived: for linear dispersion we set tight constraints placing the QG energy scale above the Planck mass; a quadratic leading LIV effect is also constrained.
}


\begin{document}
\maketitle

\newdimen\origiwspc
\origiwspc=\fontdimen2{Introduction/Motivation}
Several Quantum Gravity (QG) theories allow for a violation of the Lorentz invariance (LIV), that can manifest as a dependence of the velocity of light in vacuum on its frequency (see \cite{bib:models1,bib:models2} for examples of such models). These QG effects causing LIV are supposed to arise at a specific energy scale, namely $\mathrm{E}_{\rm QG}$, believed to be of the order of the Planck energy (\mbox{E$_{\rm Planck} \simeq 1.22 \times 10^{19}$ GeV}). If such a dependence exists, then photons of different energies emitted together by a distant source will arrive on Earth at different times. Amelino-Camelia \textit{et al.} \cite{bib:Amelino-Camelia} have parametrized the possible LIV time delays using a series expansion of powers of the photon energy over E$_{\rm QG}$. Specifically, the degree of dispersion due to LIV effects $\tau_n$ $-$ defined for two photons as the ratio of their delay in arrival ${\rm\Delta} t$ over the difference of their energies (resp. of their squared energies) ${\rm\Delta} (E^n)$ $-$ is connected to $\mathrm{E}_{\rm QG}$, to the distance of the source $\kappa_n$, and to the Hubble constant H$_0$, $n = 1$ (resp. $n = 2$) accounting for linear (quadratic) LIV effects:
\begin{eqnarray}
\tau_n = \frac{{\rm\Delta} t}{{\rm\Delta} (E^n)} \simeq s_{\pm}\frac{(1+n)}{2\mathrm{E}_{\rm QG}^n \mathrm{H}_\mathrm{0}} \kappa_n
\label{eq:tau}
\end{eqnarray}
\begin{eqnarray}
\textrm{with \quad} \kappa_n =  \int_0^z \frac{(1+z')^n\,dz'}{\sqrt{\Omega_m(1+z')^3 + \Omega_{\Lambda}}}
\end{eqnarray}
and \mbox{$s_{\pm}$ = $-$1} (resp. +1) in superluminal (subluminal) case, \textit{i.e.} if the photons are faster with larger (smaller) energies.
The cosmological parameters are fixed here to those derived from WMAP results 
\cite{bib:WMAP}
: \mbox{$\Omega_m = 0.24 \pm 0.02$}, \mbox{$\Omega_\Lambda = 0.73 \pm 0.03$}, \mbox{H$_0 = 70.4 \pm 1.4$ km.s$^{-1}$.Mpc$^{-1}$}.

A direct consequence of equation (1) is that any measurement (or limits) on the dispersion parameter $\tau_n$  with photons detected in a well-located (\textit{i.e.} with known redshift) gamma-ray source can lead to a measurement (or limits) on the energy scale $\mathrm{E}_{\rm QG}$.

The full analysis and results have been already published \cite{bib:article}, here we give an overview.

\section{Data}
\begin{figure}[t]
  \centering
  \includegraphics[width=0.442\textwidth]{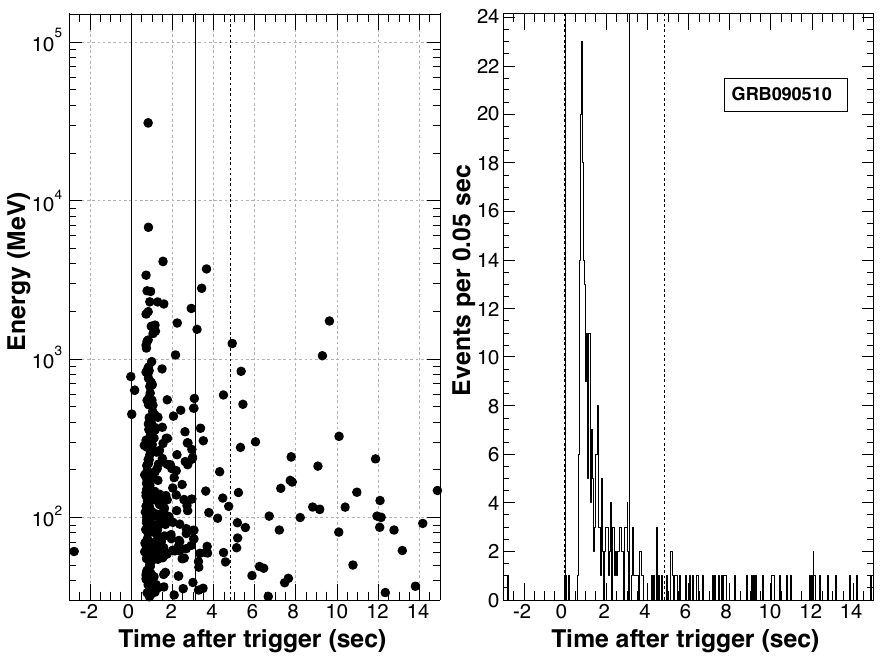}
\vspace{-3mm}
  \caption{Event energy versus event time scatter plot (left) and lightcurve (right) of the detected events from GRB090510. The external pair of vertical dashed lines shows the time interval for analysis.}
  \label{fig1}
\vspace{-1mm}
 \end{figure}

High-energy (GeV) transient emissions from distant astrophysical sources such as Gamma-ray Bursts (GRBs) and Active Galaxy Nuclei can be used to search for and constrain LIV. The \textit{Fermi} LAT and GBM collaborations have previously analyzed two GRBs in order to put such constraints.
In this analysis, four bright \textit{Fermi}-LAT GBRs are used; their distances are given in table \ref{tab:distances}. For GRB090510, the time and energy profiles are given in figure \ref{fig1}.
\begin{table}[t]
\begin{center}
\scalebox{0.9}{
\begin{tabular}{cccc}
GRB & Redshift & $\kappa_1$ & $\kappa_2$   \\ \hline
  080916C & 4.35 $\pm$ 0.15 & 4.44 & 13.50   \\
  090510 & 0.903 $\pm$ 0.003& 1.03&  1.50  \\
  090902B & 1.822 $\pm$ 0.001 & 2.07& 3.96    \\
  090926A & 2.1071 $\pm$ 0.0001& 2.37 & 4.85    \\
\end{tabular}
}
\vspace{-1.5mm}
\caption{Distances of analyzed GRBs.}
\label{tab:distances}
\end{center}
\vspace{-7mm}
\end{table}

\vspace{-4mm} 
\paragraph{Selection}
The event selection \textsc{P7\_Transient\_v6} is a less restrictive selection of LAT events optimized for signal-limited analyses such as this one. We analyzed photons with energies above 30 MeV. 
To minimize systematic errors due to GRB spectral evolution (a well known property of GRB emissions), the time intervals were chosen to focus on the main peaks of emission.
Other method-specific cuts are discussed in the next section.

\vspace{-4mm}
\section{Data analysis: methods}
Three methods, complementary in terms of sensitivity and probing different aspects of the lightcurves, have been used: PairView (PV), Sharpness Maximization Method (SMM) and Maximum likelihood (ML). Configuration details for each method are given in table \ref{tab:config_details}.
 \begin{table*}[!t]
 \begin{center}
\resizebox{0.9\textwidth}{!}{
\begin{tabular}{c|cc|cc|cc|ccccc}
GRB &  \multicolumn{2}{c}{Time Range (s)} & \multicolumn{2}{c}{$\rho$} & \multicolumn{2}{c}{\rm $N_{\rm 100}$}&$\gamma$ & \multicolumn{1}{c}{$N_{\rm template}$} & \multicolumn{2}{c}{$N_{\rm fit}$} & $E_\mathrm{\rm cut}$ (MeV)\\
         & \multicolumn{2}{c}{All Methods} & \multicolumn{2}{c}{SMM} &\multicolumn{2}{c}{PV \& SMM}&\multicolumn{5}{c}{Likelihood} \\
 & $n=1$ & $n=2$ &$n=1$ &$n=2$ &$n=1$ &$n=2$ & $n=\{1,2\}$&$n=\{1,2\}$ & $n=1$ & $n=2$ &$n=\{1,2\}$\\ \hline

  080916C & 3.53--7.89 &3.53--7.80  &  50 & 30&59 &59 &2.2 & 82 & 59 &59 &100\\
  090510 & -0.01--3.11 &-0.01--4.82 & 50 & 70&157 &168 &1.5 &148 &118 &125 &150 \\
  090902B & 5.79--14.22 &5.79--14.21 & 80 &80 &111 &111 &1.9 & 57&87 & 87&150 \\
  090926A & 8.92--10.77 &9.3--10.76&  25 & 30&60 &58 &2.2 & 53 &48 & 47&120 \\
\end{tabular}
}
  \caption{Configuration details. $\rho$ is the tuning parameter used in the SMM's sharpness measure; $N_{\rm 100}$, the number of events above 100 MeV used with PV and SMM; $\gamma$, the photon index of the spectrum of detected events assuming the energy distribution follows a power-law ($\propto e^{-\gamma}$). $E_{\rm cut}$ is the separating energy between the $N_{\rm template}$ low-energy events used for building the lightcurve template and the $N_{\rm fit}$ high-energy events used in the calculation of the likelihood.}
  \label{tab:config_details}
 \end{center}
 \end{table*}

\vspace{-2mm} 
\subsection{PairView}
This new method has been developed for this study; the principle is to calculate the following ratios between all the pairs of photons ($i,j$) with $i \neq j$:
\begin{eqnarray}
 L_{i, j}(n) = \frac{t_i-t_j}{E_i^n-E_j^n}
\end{eqnarray}

with $t_{\{i,j\}}$ and $E_{\{i,j\}}$ the arrival time and the reconstructed energy of the photon ${\{i,j\}}$, $n$ the order of LIV effect (linear for n = 1 and quad. for n = 2).
The most probable value is kept as the best estimate of the parameter $\tau_n$. This is done by using a Kernel Density estimation \cite{bib:KDE} of the distribution of $L_{i,j}$ values.

Energies below 100 MeV have been discarded to minimize possible contamination by the Band component \cite{bib:Band}. There are $N_{100}$ events left for the calculation of the L$_{i, jn}(n)$; values of $N_{100}$ for each GRB are given in table \ref{tab:config_details}. 

\subsection{Sharpness Maximization Method}
Due to LIV-induced dispersion, the time distribution of the photons is spread or equivalently the sharpness of the light curve is reduced. The method consists in finding the value of the parameter $\tau_n$ that, when inversely applied on the data, will recover the sharpness of the light curve (assumed as initially maximal).

For each tested parameter $\tau_n$: the arrival times are shifted by a factor $- \tau_n \textrm{E}^n$; the resulting modified times $t'$ are then sorted from smallest to largest; the sharpness of the resulting set of times $t'_i$ is calculated.

This method is similar to DisCan by Scargle \textit{et al.} \cite{bib:Scargle}: in their work, different definitions of the "sharpness" of the time distribution have been used.
In this study, a modification of Shannon function is used, given by the following sum over the photons $i$:
\begin{eqnarray}
S(\tau_n) = \sum_{i=1}^{N-\rho} \log \left( \frac{\rho}{t'_{i+\rho}-t'_i} \right)
\label{eq:smm}
\end{eqnarray}

where $\rho$ is a parameter defined \textit{a priori} for each GRB from simulated datasets so as to maximize the sensitivity of the method. Small (resp. large) values of $\rho$ will result in the method focusing on small (large) timescales.

\subsection{Maximum likelihood} 

A model of the emission of photons at the source is built and then used to calculate the probability that the events in the data caracterized by (t$_i$, E$_i$) have been subject to dispersion by a factor $\tau_n E^n$.

Several assumptions/simplifications have been made:  

$-$ the energies are well reconstructed (no smearing); 

$-$ the energy distribution follows a power-law spectrum of index $\gamma$ (cf. table \ref{tab:config_details}); 

$-$ no spectral variability (\textit{e.g.} variation of index $\gamma$ with time) shows up; 

$-$ the emission lightcurve (at the source) is approximated by the lightcurve of the lowest-energy events (in the data).

These assumptions lead to the following probability density function (PDF) of emission:
\begin{eqnarray}
P(t, E |\tau_n) =\frac{1}{N_{pred}} \Lambda(E) f(t-\tau_n E^n)
\label{eq:ml}
\end{eqnarray}

\vspace{-2mm}
where $N_{pred}$ is the number of photons emitted by the source,  $\Lambda(E)$ the observed power-law spectrum, $f$ a parametrization of the emission lightcurve, obtained from a 2- or 3-gaussian fit of the low energy event times. 

 A separating energy $E_{\rm cut}$ is chosen so as to split the dataset into two sets: the $N_{\rm template}$ lowest-energy events (with $E < E_{\rm cut}$) are used to build a template lightcurve, while the $N_{\rm fit}$ highest-energy events ($E > E_{\rm cut}$) are processed in the calculation of the likelihood.

\vspace{-2mm}
\section{Data analysis: confidence intervals}
The measurement of the dispersion $\tau_n$ (directly in the data) doesn't distinguish between dispersion arising from LIV effect itself (we call it $\tau_{\rm LIV}$) and spectral variability as the source, that could mimic a dispersion factor ($\tau_{\rm int}$). Hence we have

\vspace{-7mm}
\begin{eqnarray}
\tau_n = \tau_{\rm LIV} + \tau_{\rm int}
\end{eqnarray}
Previous studies have ignored the term $\tau_{\rm int}$, assuming \mbox{$\tau_n = \tau_{\rm LIV}$}. Here, we calculate a first confidence interval (CI) on the total degree of dispersion $\tau_n$; we then give a CI for the dispersion possibly arising because of LIV effects $\tau_{\rm LIV}$, based on conservative assumptions on $\tau_{\rm int}$. 
\subsection{Confidence intervals on $\tau_n$}
The CI is done in differents ways for PV/SMM and ML.

\vspace{1mm}
For PV and SMM, sets are created by randomizing the associations of time and energy from the original dataset. By definition, the time and energy distributions will remain the same; yet the mixing of times and energies is expected to remove any possible dispersion in the data. A hundred thousand of these sets are produced, the PV or SMM method is applied to get the best estimate $\hat{\tau}_n$ of the dispersion parameter $\tau_n$ for each set. The resulting distribution, $f_r$, is used to approximate the PDF of the measurement error on $\tau_n$ (general case of any $\tau_n$): $\epsilon = \hat{\tau}_n - \tau_n$. We then calculate a CI for $\tau_n$ from the quantiles of $f_r$.

\vspace{1mm}
For ML, calibrated CIs were calculated using Monte Carlo simulated sets.
These sets have the same statistics, lightcurve model and spectrum as the original data. No intrinsic dispersion was artificially added. Each simulated data set produces a lower limit and an upper limit on $\tau_n$ from cuts on the likelihood profile \footnote{the value of the cut is set so as to get exact coverage}.
The calibrated lower (resp. upper) limit of the CI is taken as the mean of the distribution of the per-set individual lower (upper) limits.

\subsection{Confidence intervals on $\tau_{\rm LIV}$}
\fontdimen2\font=0.49ex
No reliable model of GRB emission at LAT energies (\mbox{$E>100$ MeV}) has been produced yet, therefore a conservative estimation of the impact of $\tau_{\rm int}$ has been chosen. We assume that the measurements of $\tau_n$ are dominated by GRB intrinsic effects, so the PDF of $\tau_{\rm int}$ is to match the dispersion allowed by the data:  $\langle\tau_{\rm int}\rangle = 0$, since all the measures on $\tau_n$ don't exclude a trivial dispersion; and the width has to match the width of $\tau_n$. 
Since the intrinsic effect is set to perfectly match any observed dispersion, the CIs on $\tau_{\rm LIV}$ are built so that they have the largest possible width. 
The way the PDF $P_{\tau_{\rm int}}(\hat{\tau}_{\rm int})$ is implemented depends on the method.
\fontdimen2\font=\origiwspc

\vspace{0.8mm}
For PV and SMM, the  PDF of the measurement error on $\varepsilon ' = \hat{\tau}_n - \tau_{\rm LIV}$ is built 
as the autocorrelation of $f_r$ (from distribution of the best estimates calculated on the randomized sets, cf. section 4.1) with argument $\varepsilon '$. By definition, the resulting PDF is symmetric; its quantiles are used to calculate a CI for $\tau_{\rm LIV}$.

\vspace{0.8mm}
For ML, the model of emission in equation~\ref{eq:ml} is changed: 
\begin{eqnarray}
P(E,t|\tau_{\rm LIV};\tilde{\tau}_{\rm int})= \frac{1}{N_{pred}} \Lambda(E) f(t-\tau_{\rm LIV} E^n -\tilde{\tau}_{\rm int} E^n)
\end{eqnarray}
\fontdimen2\font=0.51ex
For each simulated set, a random value of $\tilde{\tau}_{\rm int}$ is drawn. The calculation of $\tau_{\rm LIV}$ is then done with the ML method; the means of the LL and UL distribution provide the CI for $\tau_{\rm LIV}$.
\fontdimen2\font=\origiwspc

\vspace{0.8mm}
\fontdimen2\font=0.51ex
We end up with symmetric CIs on $\tau_{\rm LIV}$, that correspond to a worst case (yet reasonable) scenario for GRB-intrinsic effects. These CIs on $\tau_{\rm LIV}$ are less stringent, though far more robust with respect to the presence of GRB-intrinsic effects.
\fontdimen2\font=\origiwspc

\vspace{-4mm}
\section{Results}
\paragraph{Total degree of dispersion $\tau_n$}

\begin{figure*}[!t]
\vspace{-1mm}
\centering
\resizebox{\textwidth}{!}{
\includegraphics[width=0.264\textwidth]{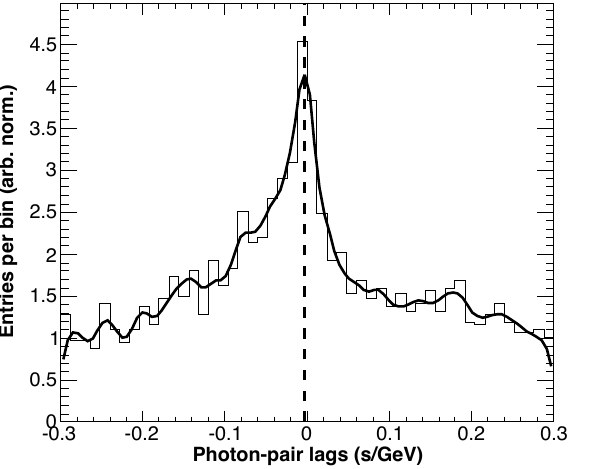}
\includegraphics[width=0.264\textwidth]{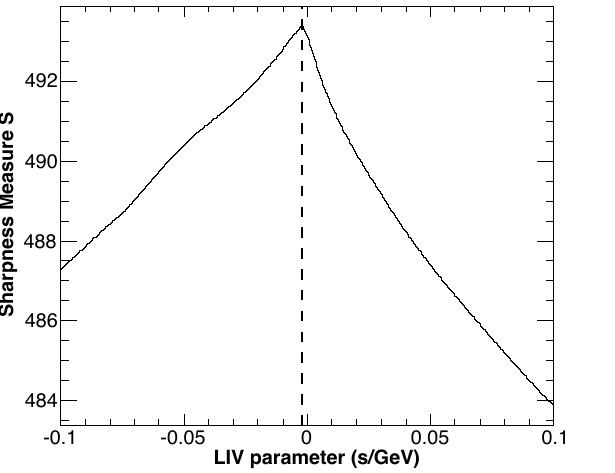}
\includegraphics[width=0.248\textwidth]{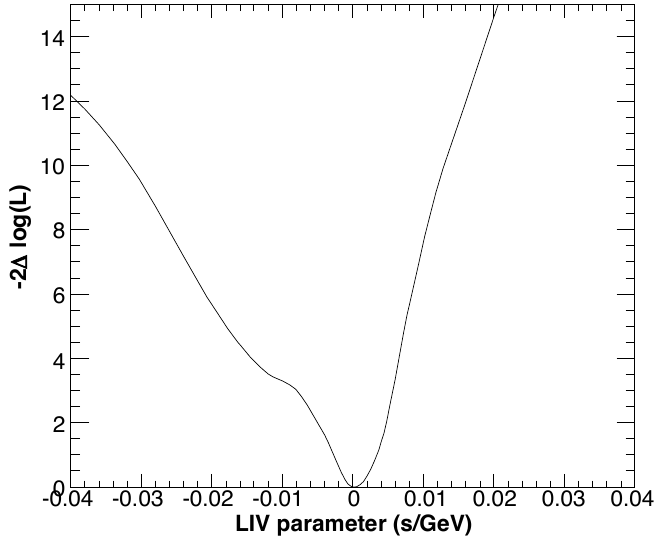}
}
\vspace{-8mm}
\caption{Application of the PV, SMM and ML methods on GRB090510 for the linear case (\mbox{n = 1}).}
\label{fig2}
\end{figure*}

Figures
\ref{fig2} demonstrate the application of the PV, SMM and ML methods on GRB090510 for the linear case (n=1).

The measurements of $\bf \tau_{\bf n}$ parameter for each method, for both linear and quadratic cases, are summarized in \mbox{table \ref{tab:tau_n}}. The limits obtained on the four bursts with the three different methods are compatible with no dispersion. The likelihood method gives the more stringent constraints. 

\begin{table}[!t]
\begin{flushright}
 
\resizebox{1.05\columnwidth}{!}{
\begin{tabular}{l|ccc|ccc|ccc}
 GRB Name &   \multicolumn{3}{c}{PairView} & \multicolumn{3}{c}{SMM}  & \multicolumn{3}{c}{Likelihood}\\
&\multicolumn{9}{c}{(Lower Limit, Best Value, Upper Limit) (s\,GeV$^{-1}$) $n=1$}\\ 
\hline
080916C   &-0.46 & 0.69 & 1.9  &-0.49 & 0.79 & 2.3  &-0.85 & 0.1 & 0.77  \\
090510   ($\times 10^{3}$)&-73 & -14 & 27  &-74 & -12 & 30   &-9.8 & 1 & 8.6  \\
090902B   &-0.36 & 0.17 & 0.53  &-0.25 & 0.21 & 0.62   &-0.63 & 0.25 & 0.96  \\
090926A   &-0.45 & -0.17 & 0.15  &-0.66 & -0.2 & 0.23  &-0.56 & -0.18 & 0.18  \\
&\multicolumn{9}{c}{(Lower Limit, Best Value, Upper Limit) (s\,GeV$^{-2}$) $n=2$}\\ 
\hline
080916C   &-0.18 & 0.45 & 1.1  &-0.0031 & 0.88 & 2  &-0.83 & 0.12 & 0.8  \\
090510   ($\times 10^{3}$)&-3.9 & -0.63 & 0.88  &-4.1 & -0.68 & 0.85  &-0.32 & -0.1 & 0.23  \\
090902B   ($\times 10^{3}$)&-26 & 17 & 48  &-18 & 24 & 60   &-120 & 10 & 110  \\
090926A   &-0.18 & -0.021 & 0.13  &-0.12 & -0.06 & 0.012  &-0.44 & -0.06 & 0.14  \\
\end{tabular}
}

  \caption{Measurements of the total degree of dispersion $\tau_n$ for each method. Limits are for a 99\% double-sided confidence level.
  }
  \label{tab:tau_n}
 \end{flushright}
 \vspace{-4mm}
 \end{table}

\vspace{-4mm}
\paragraph{LIV-induced dispersion  $\tau_{\rm \bf LIV}$}
New CIs on the LIV-induced dispersion parameter $\tau_{\rm LIV}$ have been produced, as explained in section 4. We don't provide here their values for each GRB; limits on E$_{\rm QG}$ using these specific CIs have been produced (see below).

\begin{figure*}[!t]
\centering

\includegraphics[width=0.448\textwidth]{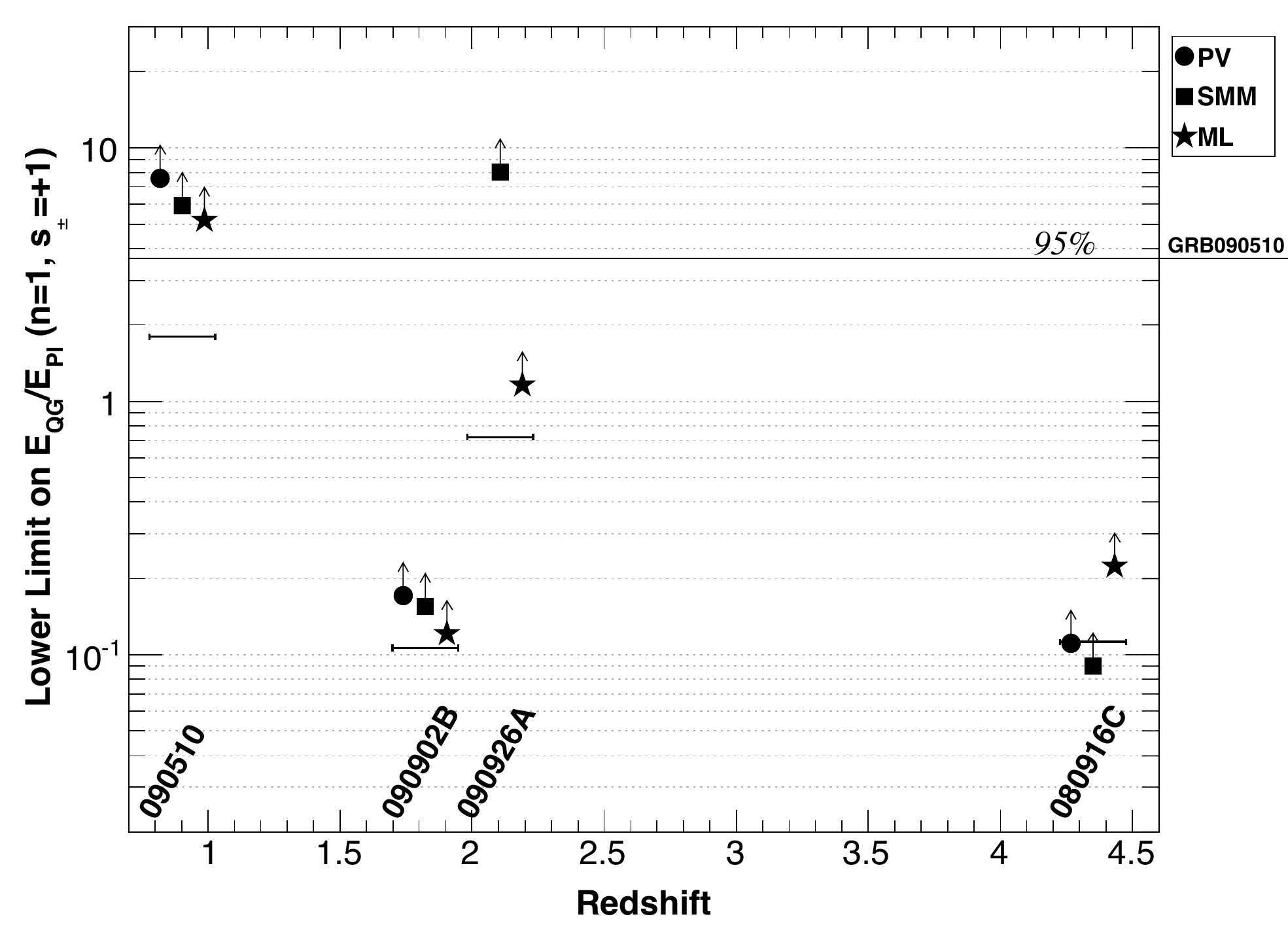}
\includegraphics[width=0.44\textwidth]{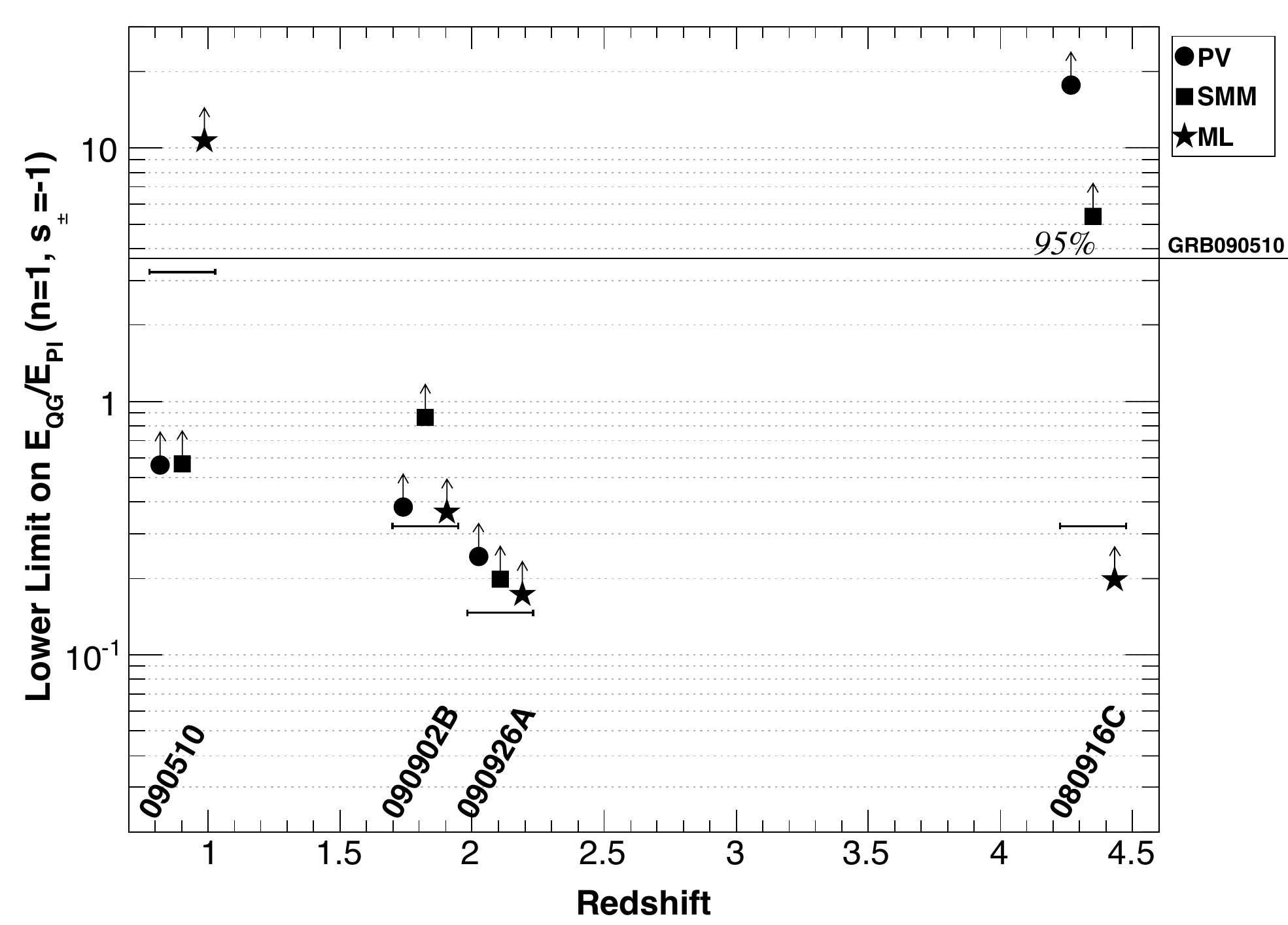}\\
\includegraphics[width=0.448\textwidth]{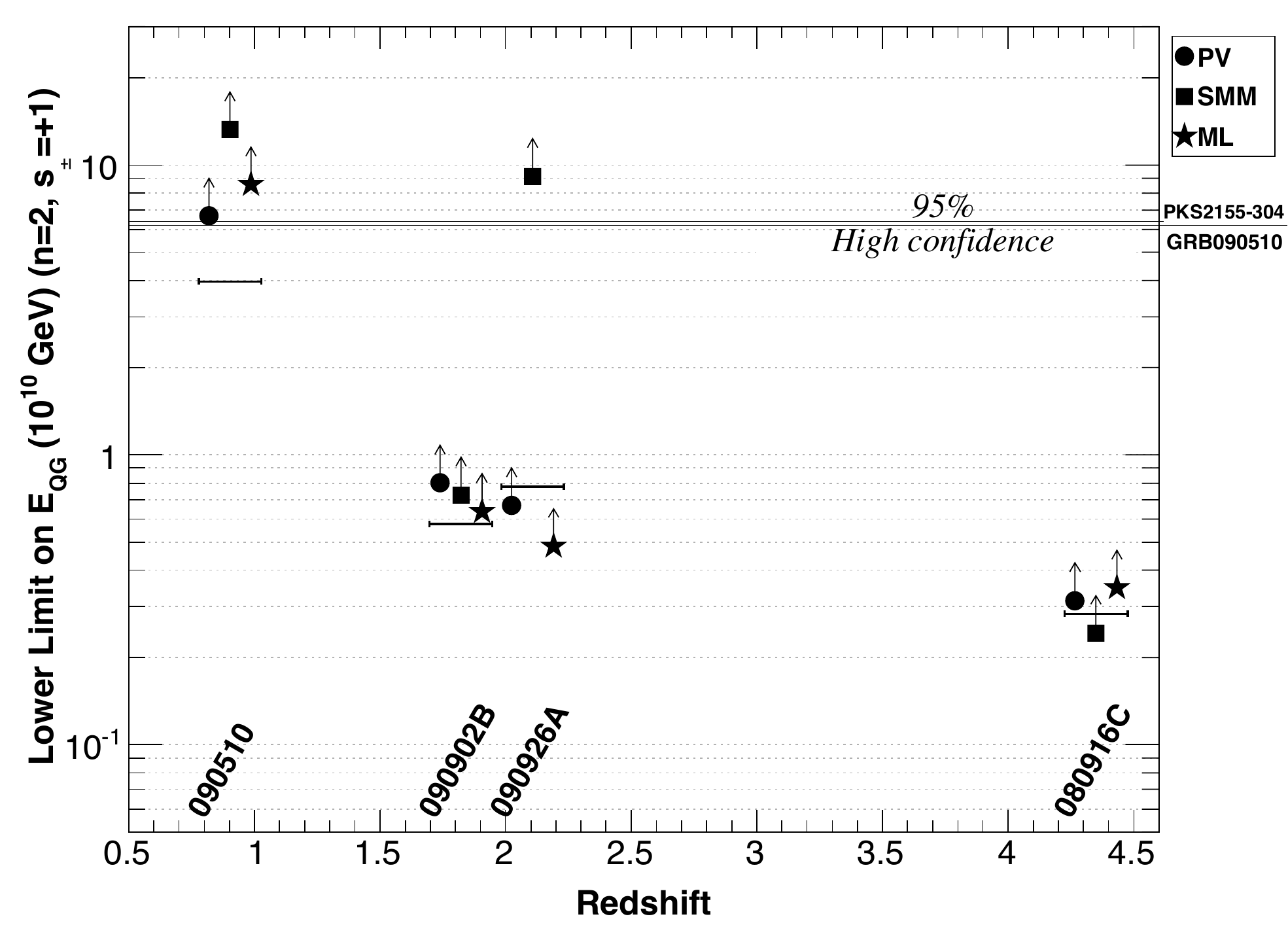}
\includegraphics[width=0.44\textwidth]{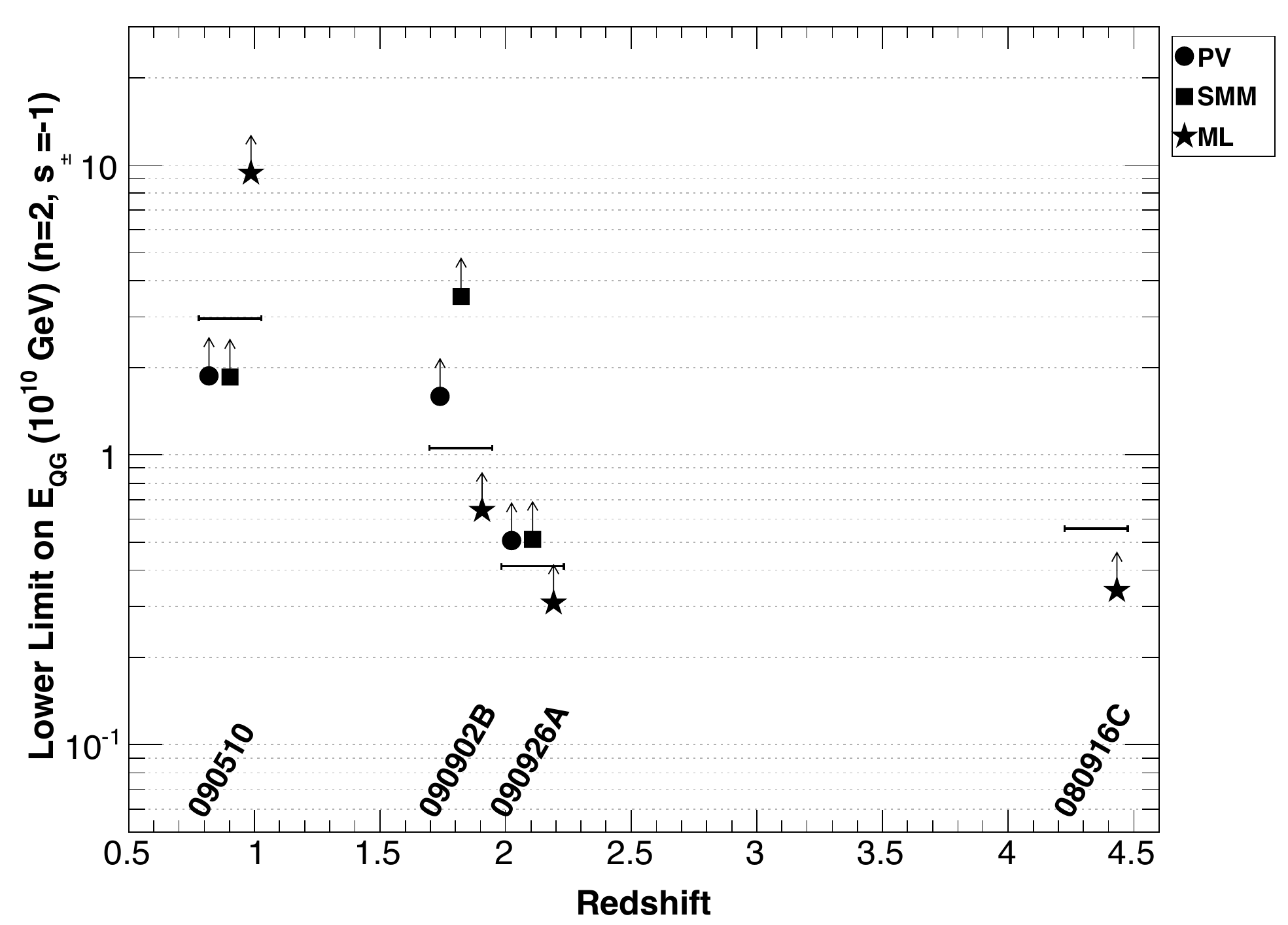}
\vspace{-4mm}

\caption{95\% one-sided CL limits on E$_{\rm QG}$ for subluminal (\mbox{s = +1}, left) and superluminal (\mbox{s = -1}, right) cases, for a linear (\mbox{n = 1}, top) and quadratic (\mbox{n = 2}, bottom) LIV effect. Each triplet of points corresponds to one GRB and shows, left to right, the limits obtained with PV, SMM and ML (from $\tau_{n}$). The horizontal bars correspond to an averaged over the three methods limit for the intrisic-corrected case (from $\tau_{\rm LIV}$). 
}
\label{fig3}
\vspace{-4mm}
\end{figure*}

\vspace{-2mm}
\paragraph{QG energy scale E$_{\rm \bf QG}$}
It is possible to use the limits on the total dispersion $\tau_n$ to set limits on the energy scale E$_{\rm QG}$ at which QG effects are expected to become important, using the equation \ref{eq:tau}.
Lower limits on E$_{\rm QG}$ for are plotted in figure \ref{fig3}. The best limit for subluminal case is obtained with GRB090510 and corresponds to 
$\textrm{E}_{\rm QG,1}>7.6 \textrm{ E}_{\rm Planck}$  and $\textrm{E}_{\rm QG,2}>1.3\times 10^{11} \textrm{ GeV}$ for a linear and a quadratic LIV effect, respectively. 

This figure also shows E$_{\rm QG}$ limits derived with data from the LIV-induced dispersion parameter $\tau_{\rm LIV}$, after accounting for intrinsic effects for each GRB (horizontal bars).
These limits are less stringent, yet they are much more robust with respect to the presence of GRB-intrinsic dispersion than can masquerade as dispersion induced by LIV. The best limit for linear/subluminal case is obtained with GRB090510: $\textrm{E}_{\rm QG,1} \gtrsim 2 \textrm{ E}_{\rm Planck}$.

Horizontal lines show current most constraining limits (not accounting for intrinsic effects) obtained with GRB 090510 (\textit{Fermi} LAT/GBM \cite{bib:GRB090510}) and PKS 2155-304 (\mbox{H.E.S.S. \cite{bib:PKS2155}}). We improve these limits by a factor 2 to 4.

\vspace{-2mm}
\section{Systematic uncertainties}
\paragraph{GRB-intrinsic effects}

Previous analyses of \textit{Fermi} \mbox{GRB} data have shown that the emission falls into different spectral components: a Band component at MeV energies, and a power-law component at higher energies. The difference in time emission between the two components could be misidentified as a LIV effect. The cuts (E $>100$ MeV) used for PV and SMM limit the contribution of Band-originated events. The cuts at a lower energy (E $>30$ MeV) used for the ML method makes it possibly more sensitive to this effect. However, a dedicated analysis of the data didn't show any sign of time lag between the two spectral components.

Also, spectral evolution has been detected in many \textit{Fermi}-LAT GRBs and can be misinterpreted as LIV effect.  Focusing on the time windows with higher variability (namely the peaks of the lightcurve) allow to reduce the influence of this effect in present analysis. 

\vspace{-3.7mm}
\paragraph{Instrument}
The energy reconstruction uncertainty due to high off-axis angles is negligible. Difference between true and reconstructed energies leads to systematics of  $\sim$ 10\% (n = 1) and $\sim$ 15\% (n = 2). For the ML method, the dependence of the effective area on the energy has been neglected: this systematic uncertainty is dominated by statistical uncertainty on the spectrum.

\vspace{-3.7mm}
\paragraph{Other effects}
Background contamination is negligible (very low rate). Uncertainty on the redshift is $\sim$ 1\% (GRB080916C) and $\sim$ 0.1\% (other). The uncertainty on the cosmological parameters is $\sim$ 3\%.

\section{Conclusions}
The analysis of the four GRBs  did not lead to the observation of a significant energy dependence of the speed of light in vacuum. However, it was possible to place robust limits on the E$_{\rm QG}$ scale. The best limits were obtained for the shortest GRB (GRB090510), which is also the closest one: for the linear/subluminal case, E$_{\rm QG, 1}>$ 7.6 E$_{\rm Planck}$ using the total degree of dispersion $\tau_n$ and E$_{\rm QG, 1} \gtrsim$ 2 E$_{\rm Planck}$ using the LIV-induced degree of dispersion $\tau_{\rm LIV}$ (limits at \mbox{95\% CL}). Observing short GRBs at high redshift would possibly improve this limit. A discussion of the systematics, including taking into account possible source effects, has also been conducted. 
Detailed analysis and discussion can be found in \cite{bib:article}.

\footnotesize  
\fontdimen2\font=0.46ex
The $Fermi$ LAT Collaboration acknowledges support from a number of agencies and institutes for both development and the operation of the LAT as well as scientific data analysis. These include NASA and DOE in the United States, CEA/Irfu and IN2P3/CNRS in France, ASI and INFN in Italy, MEXT, KEK, and JAXA in Japan, and the K.~A.~Wallenberg Foundation, the Swedish Research Council and the National Space Board in Sweden. Additional support from INAF in Italy and CNES in France for science analysis during the operations phase is also gratefully acknowledged.
\fontdimen2\font=\origiwspc

\normalsize


\begin{thebibliography}{}
\vspace{-1mm}
\bibitem{bib:models1} Mattingly, Living Rev.Rel. 8 (2005)
\bibitem{bib:models2} Jacobson \textit{et al.}, Annals Phys. 321 (2006) 150-196
\bibitem{bib:Amelino-Camelia} Amelino-Camelia \textit{et al.}, Nature 393 (1998) 763-765
\bibitem{bib:WMAP} Komatsu \textit{et al.}, Astrophys.J.Suppl. 192 (2011) 18 
\bibitem{bib:article} Vasileiou \textit{et al.}, Phys. Rev. D 87 (2013) 122001 \\{\small arXiv:1305.3463}
\bibitem{bib:KDE} Cranmer, Comp. Phys. Com.  1363 (2001) 198-207
\bibitem{bib:Band} Band \textit{et al.}, Astrophys.J. 413 (1993) 281-292 
\bibitem{bib:Scargle} Scargle \textit{et al.}, Astrophys.J. 673  (2008) 972 
\bibitem{bib:GRB090510} Abdo \textit{et al.}, Nature 462 (2009) 331
\bibitem{bib:PKS2155} Abramowski \textit{et al.}, Astropart. Phys. 34 (2011) 738
\end{thebibliography}
\end{document}